\documentclass{emulateapj}

\newcommand{\HM}{{\rm H^-}}
\newcommand{\HH}{{\rm H_2}}
\newcommand{\lsim}{\lesssim}
\newcommand{\gsim}{\gtrsim}

\begin{document}

\title{Feedback from first radiation sources: $\HM$ photodissociation}
\author{Leonid Chuzhoy$^{1}$,  Michael Kuhlen$^{2}$ and Paul R. Shapiro$^{1}$  }
\bigskip

\altaffiltext{1}{McDonald Observatory and Department of Astronomy,
The University of Texas at Austin, RLM 16.206, Austin, TX 78712,
USA; chuzhoy@astro.as.utexas.edu}

\altaffiltext{2}{Institute for Advanced Study, Princeton, NJ, 08540,
USA; mqk@ias.edu}

\begin{abstract}
During the epoch of reionization, the formation of radiation sources
is accompanied by the growth of a $\HM$ photodissociating flux. We
estimate the impact of this flux on the formation of molecular
hydrogen and cooling in the first galaxies, assuming different types
of radiation sources (e.g. Pop II and Pop III stars, miniquasars).
We find that $\HM$ photodissociation reduces the formation of $\HH$
molecules by a factor of $F_{\rm s}\sim 1+10^3 k_s xf_{\rm
esc}^{-1}\delta^{-1}$, where $x$ is the mean ionized fraction in the
IGM, $f_{\rm esc}$ is the fraction of ionizing photons that escape
from their progenitor halos, $\delta$ is the local gas overdensity
and $k_s$ is an order unity constant which depends on the type of
radiation source. By the time a significant fraction of the universe
becomes ionized, $\HM$ photodissociation may significantly reduce
the $\HH$ abundance and, with it, the primordial star formation
rate, delaying the progress of reionization.

\end{abstract}


\section{\label{Int}Introduction}
The first stars in the $\Lambda$CDM universe are believed to have
formed inside dark-matter-dominated minihalos filled with mostly
neutral, metal-free gas of virial temperature $T_{\rm vir} < 10^4$
K, when $\HH$ molecules formed in sufficient abundance to cool the
gas radiatively to $\sim 10^2$ K. If, as currently thought, these
stars were massive, hot, and luminous, they may have contributed
significantly to the reionization of the universe, which CMB
polarization observations by WMAP indicate was highly ionized by
$z\sim 10$ \citep{Sp}.  The release of ionizing UV radiation by
minihalos and other sources (e.g. stars in more massive halos, with
$T_{\rm vir} > 10^4$ K, or miniquasars), required to explain
reionization, must have been accompanied by radiation release at
energies below the H Lyman limit, as well, however.  This may, in
turn, have limited the $\HH$ abundance inside minihalos and their
ability to form stars, thereby limiting their contribution to cosmic
reionization.

In the absence of dust and at densities below the three-body formation
regime ($n \lesssim 10^{10}$ cm$^{-3}$), the most important reaction
for the production of $\HH$ is
\begin{equation}
\label{fH2}
{\rm H^{-}+H\rightarrow H_2 + e^-},
\end{equation}
(e.g., Shapiro \& Kang 1987 and refs. therein) with reaction rate
$k_{-}=1.3\times 10^{-9}\;{\rm cm^{3}\; s^{-1}}$ \citep{Schm}. Once
formed, $\HH$ can be destroyed by collisions with other species
\begin{eqnarray}
{\rm H_{2}+H^+\rightarrow H_2^+ + H}, \\
{\rm H_{2}+H\rightarrow H + H+H}, \\
{\rm H_{2}+e^-\rightarrow H + H+ e^-},
\end{eqnarray}
or by photodissociation via Lyman-Werner band photon absorption
\begin{eqnarray}
\label{fdes}
{\rm H_2+\gamma \rightarrow H + H}.
\end{eqnarray}
The latter process becomes dominant once a substantial UV background
is built up between 912 and 1110 $\AA$, providing a feedback mechanism
against the formation of new radiation sources
\citep{Hai97,Hai00,Ciar,Mach,Mes}.

In this paper we explore the impact of another feedback mechanism, the
photodissociation of $\HM$,
\begin{equation}
{\rm H^{-}+\gamma\rightarrow H + e^-}.
\end{equation}
The cross-section for photodissociation of $\HM$  is well
fitted by \citep{Wish}
\begin{equation}
\sigma_{-}(\epsilon)=2.1\times 10^{-16} \frac{(\epsilon -0.75)^{3/2}}{\epsilon^{3.11}} \; {\rm cm}^2,
\end{equation}
where $\epsilon$ is the photon energy in eV. The cross section is
zero below a threshold of $\epsilon < 0.755$eV, the binding energy
of the second electron. In the absence of the UV background, the
primary mode of $\HM$ destruction is the formation of $\HH$ (Eq.
[\ref{fH2}]), \footnote{When gas fractional ionization is high
($x\gsim 0.01$) mutual neutralization with ${\rm H^+}$ can provide
another efficient channel for $\HM$ destruction. However, typically
the fractional ionization of minihalos is much lower.} so
introducing the $\HM$ photodissociating flux reduces the $\HH$
formation rate by a factor
\begin{equation}
\label{sup}
F_{\rm s}=1+\frac{\zeta_{-}}{k_{-}n_{\rm H}},
\end{equation}
 where $\zeta_{-}=\int n_\gamma(\epsilon)\sigma_{-}(\epsilon)c d\epsilon$
%\begin{equation}
%\zeta_{-}=\int n_\gamma(\epsilon)\sigma_{-}(\epsilon)c d\epsilon,
%\end{equation}
is the photodissociation rate per $\HM$ ion, $n_{\rm H}$ is the
hydrogen atom number density and $n_\gamma(\epsilon)$ is the number
density of photons with energy $\epsilon$.\footnote{This
approximation for $F_{\rm s}$ breaks down when its value exceeds
$\sim 50 $, since for such UV intensities ${\rm H^{+} +H\rightarrow
H_2^+ + \gamma}$ reaction becomes a dominant channel of $\HH$
production (assuming reaction rates given by \citet{SK}). Note also
that $k_{-}$ is still uncertain to within a factor of a few (see
Glover et al. 2006), and this uncertainty carries over to $F_{\rm
s}$ when $F_{\rm s} \gg 1$.}  Hence the importance of this mechanism
depends primarily on the local density ratio of $\HM$
photodissociating photons and hydrogen atoms.

The impact of $\HM$ photodissociation differs from that of ${\rm
H_2}$ by two fundamental characteristics. First, the time required
for $\HM$ abundance to approach equilibrium is very short (typically
less than 10000 years), while for $\HH$ the equilibration time can
exceed the Hubble time. Therefore, when gas is exposed to a
\textit{transient} UV flux, produced by nearby Pop III stars, for
example, $\HM$ photodissociation can generally be ignored, as it
does not affect the subsequent thermal and chemical evolution.
Secondly, photons that make up the $\HH$ photodissociating
background are destroyed after a few percent of the Hubble time, as
they redshift into one of the Lyman series resonances, and must be
replenished continuously. By contrast, photons that constitute the
$\HM$ photodissociating background are very rarely destroyed, which
allows them to accumulate over time. Consequently the importance of
$\HM$ photodissociation increases over time, and as we show in this
paper, by the time a significant ($\sim 10$ \%) fraction of the
Universe is ionized, ${\rm H^-}$ photodissociation may result in a
drastic reduction of the molecular hydrogen abundance. This in turn
may lead to a reduced star formation rate and delay the progress of
reionization.

Recently, \citet{Gl} considered the suppression of $\HH$ formation
due to the photodissociation of $\HM$ and $\rm H_2^+$. Whereas
\citet{Gl} focused on the local feedback around and inside HII
regions created by Pop III stars, we treat the problem globally and
also consider long range effects due to the much lower optical depth
of the universe below the Lyman limit.

The paper is organized as following. In \S 2 and 3, we estimate the
intensity of $\HM$ photodissociating flux produced by UV and X-ray
sources, respectively. In \S 4, we discuss the implication of our
results for gas cooling in minihalos.

\section{$\HM$ photodissociating background - UV sources}
\subsection{Recombination products}

Since the first radiation sources are expected to form within
overdense gas clouds, only the escaping fraction of their ionizing
photons, $f_{\rm esc}$, was available for ionization of the diffuse
IGM. The rest was absorbed within the host halos and, via the
process of radiative recombination, converted into lower energy UV
photons. Since the universe during that epoch is transparent to most
non-ionizing UV photons, \footnote{An exception occurs for photons
whose frequency is close to one of the high ($n>2$) Lyman
resonances, which, following their absorption by hydrogen atoms, are
further split into two or more lower energy photons. For Ly$\alpha$
photons, the optical depth is also very high, but in their case the
absorption in almost all cases is followed by reemission, with the
destruction probability being extremely low \citep{FP}.  Also, at
the very early stage of reionization ($x\ll 1$) the presence of
${\rm H_2}$ molecules makes the universe opaque in the Lyman-Werner
range. However, since their initial abundance ($\sim 10^{-6}$) is
already very low, the number of photons they destroy is negligible.}
almost all of them add to the $\HM$ photodissociation background.

Neglecting recombinations in the diffuse IGM, the mean ionization is
$x=N_{\rm ib}f_{\rm esc}$, where $N_{\rm ib}$ is the total number of
ionizing photons per baryon produced up to this point. Inside halos,
the recombination time is quite short, and so the number of
ionizations taking place there, $N_{\rm ib}(1-f_{\rm
esc})=x(1-f_{\rm esc})/f_{\rm esc}$, is almost equal to the number
of electron recombinations to $n\geq 2$ states (i.e., recombinations
which do not result in emission of additional ionizing photons),
$N_{\rm rec}$. Therefore the average $\HM$ photodissociating rate is
given by
\begin{equation}
\label{zet} \zeta_-=N_{\rm rec}n_{\rm bar}c\langle\sigma_-\rangle=x
n_{\rm bar}c \langle\sigma_-\rangle \left(\frac{1-f_{\rm
esc}}{f_{\rm esc}}\right),
\end{equation}
where $n_{\rm bar}$ is the mean baryon density and
$\langle\sigma_-\rangle$ is the average cross-section per
recombination photon times the average number of photons per
recombination, $\langle\sigma_-\rangle=\int
(j_\epsilon/\alpha_{rec}n_e
n_p\epsilon)\sigma_-(\epsilon)d\epsilon$. Note that since
emissivity, $j_\epsilon$, is proportional to $n_e n_p$,
$\langle\sigma_-\rangle$ is in fact independent of $n_e$ and $n_p$.
Using Osterbrock's (1989, Sec.~4.3) calculation of the recombination
spectrum, $(j_\epsilon/\alpha_{rec}n_e n_p)$, and assuming that the
temperature of the recombining gas is close to $10^4$ K, we find
$\langle\sigma_-\rangle= \;3.4\times 10^{-17} {\rm cm^2}$.

By combining equations (\ref{sup}) and (\ref{zet}), we can estimate the importance of the $\HM$ photodissociation due
 to recombination radiation.
Assuming that most of the recombinations occurred recently, we find
that, the recombination radiation alone will suppress the $\HH$
formation rate by
\begin{eqnarray}
\label{supuv}
 F_{\rm
s}=1+800\delta^{-1}\left(\frac{x}{0.1}\right) \left(\frac{f_{\rm
esc}}{0.1}\right)^{-1} (1-f_{\rm esc}),
\end{eqnarray}
where  $\delta=1.08n_{\rm H}/n_{\rm bar}$ is the local overdensity. Here we
have neglected recombinations in the diffuse intergalactic medium
(IGM) and the associated $\HM$ dissociating photons from these
recombinations, but these would only further increase $F_{\rm s}$.

\begin{figure}[t]
\resizebox{\columnwidth}{!} {\includegraphics{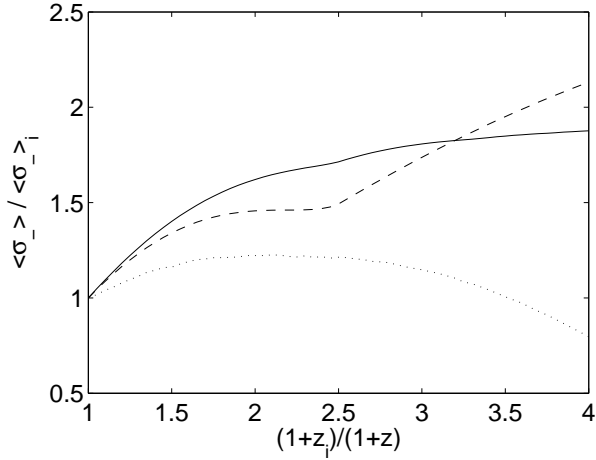}}
%\vspace*{2.5in}
\caption{Redshift evolution of the average $\HM$ photodissociation
cross-section of the UV photons produced by recombination (dotted
line), excitations by non-thermal electrons (dashed line) and
massive Pop III stars (solid line).} \label{fig:zdep}
\end{figure}

Cosmological redshift can affect the photodissociation rate by
shifting the spectrum to longer wavelengths. Initially this leads to
an increase in $\langle \sigma_- \rangle$ due to the $\epsilon^{-3/2}$
dependence of the cross-section for $\epsilon \gg 0.755$
eV. Eventually, as more and more of the spectrum is shifted below the
threshold, the cosmological redshift begins to decrease the
dissociation rate. For recombination photons this redshift effect is
small, and the transition to $\langle \sigma_- \rangle$-depression
occurs at a redshift factor of $(1+z_i)/(1+z) \approx 2.5$, see
Figure~\ref{fig:zdep}.

\subsection{Direct emission}
Unlike ionizing photons, whose intensity is heavily attenuated both
in stellar atmospheres and in their host galaxies, most of the
photons with frequencies below the Lyman limit escape freely into
the IGM. From then on, photons with frequency below Ly$\beta$
undergo no evolution apart from cosmological redshift. By contrast,
within a small fraction of the Hubble time, most photons with
frequency between Ly$\beta$ and the Lyman limit are split by cascade
into two or more photons after being redshifted into one of the
hydrogen resonances. Most of the cascade products, which include
lines such as Ly$\alpha$, H$\alpha$, and H$\beta$, as well as a
continuum spectrum produced by the two photon transition
$2s\rightarrow 1s$, are above the $0.755$ eV threshold for $\HM$
photodissociation.

\begin{figure}[t]
\resizebox{\columnwidth}{!} {\includegraphics{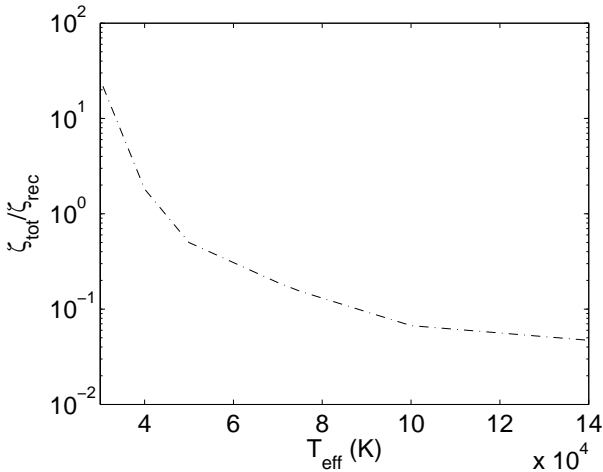}}
%\vspace*{2.5in}
\caption{The ratio between the total $\HM$ photodissociation rate
and the photodissociation by recombination products alone for star
with different effective surface temperature. }
\label{fig:zetaratio}
\end{figure}

The relative importance of these directly emitted $\HM$ dissociating
photons depends on the nature of the UV sources.
Figure~\ref{fig:zetaratio} shows the increase of the $\HM$
dissociation rate due to inclusion of direct emission from
metal-poor Pop III stars, which we calculated using the stellar
atmosphere models of \citet{Sch}. Predictably, for very massive Pop
III stars, with surface temperatures $\sim 10^5$ K, adding the
stellar continuum below the Lyman limit to the recombination
spectrum increases the photodissociation rate by only $\sim 10\%$.
If, on the other hand, most of the early ionizing flux was produced
by stars with masses below $10 M_\odot$, whose continuum emission is
stronger at lower frequencies, then the total $\HM$ dissociation
rate would be tripled at least. Likewise, direct emission may be
important if most of the UV photons were produced by miniquasars.
For example, assuming that their spectrum can be approximated by a
power law, $L_\nu\propto \nu^{-1.7}$, with a cutoff below $0.75$ eV,
adding the directly emitted photons to the recombination products
increases the total photodissociation rate by a factor of $\sim 5$.

\section{$\HM$ photodissociating background - X-ray sources}
It has been suggested that X-ray photons could contribute a large
fraction of the energy emitted by the first radiation sources
\citep[e.g.][]{RO}. By increasing the number of free electrons,
X-rays can boost the production of $\HM$, and thus of $\HH$,
providing a positive feedback to the formation of new sources
\citep{Hai00, Kuh}. This effect, however, would be at least
partially offset by an increase of the $\HM$ photodissociating
background, caused by conversion of X-rays into UV photons.

The absorption of an X-ray photon is followed by release of a
non-thermal electron, which then loses some of its energy by
inelastic collisions with atoms before it can thermalize its energy
by elastic scattering with ions and other electrons. When the gas
ionization fraction is low ($x\lsim 0.05$), the photoelectron splits
most of its energy evenly between collisional ionizations and
excitations of hydrogen atoms \citep{SvS}. Using electron-hydrogen
excitation cross-sections \citep{GSS,SKD}, we find that around $\sim
5/6$ of the excitations are to the 2p level, which are followed by
emission of a Ly$\alpha$ photon. Most of the remaining excitations
are to the 3p level, which decays via emission of one H$\alpha$
photon and a subsequent two-photon decay from the 2s level. The
Ly$\alpha$, H$\alpha$ and two-photon continuum each produce roughly
equal contributions to $\HM$ photodissociation. Per ionization, the
average intensity-weighted cross-section for these photons is
$\langle\sigma_{-}\rangle= 1.6 \times 10^{-17} {\rm cm^2}$.  Due to
the low number of UV photons produced during this phase, the
formation of $\HH$ is not strongly affected
\begin{eqnarray}
 F_{\rm s}=1+4\delta^{-1}\left(\frac{x}{0.01}\right).
\end{eqnarray}

After the ionized fraction climbs above $x\sim 0.05$, most of the
energy of the non-thermal electrons is converted to heat. However,
simultaneously with the growth of the ionized fraction, the
temperature of the gas rises, and as it crosses $10^4$ K, the
collisions between thermal electrons and atoms begin to dissipate
the energy added by X-rays, mainly via emission of Ly$\alpha$
photons. Neglecting gas clumping, we find that the number of emitted
Ly$\alpha$ photons per hydrogen atom is
\begin{equation}
\label{Na} N_\alpha=4.6\times 10^{-8} \; {\rm cm^{3}\; s^{-1}} \int
x(1-x)e^{-1.18\times 10^5/T}\;n_H\;dt .
\end{equation}
Assuming for simplicity that $x$ and $T$ are constants, we can
rewrite the equation (\ref{Na}) as
\begin{equation}
\label{Na2} N_\alpha=11.2
\left(\frac{\tau_{e,X}}{0.05}\right)\left(\frac{(1-x)e^{-1.18\times
10^5/T}}{10^{-4}}\right) ,
\end{equation}
where $\tau_{e,X}=\int xn\sigma_T dt$ is the Thompson optical depth
from the epoch of partial ionization by X-rays. If X-ray
preionization contributes at least half of the $\tau_e\sim 0.1$
measured by WMAP (i.e. $\tau_{e,X}\approx 0.05$), hydrogen atomic
de-excitations in the diffuse IGM may produce $\gsim 30$ Ly$\alpha$
photons per baryon.

The suppression of $\HH$ formation due to $\HM$ photodissociation by
Ly$\alpha$ photons is
\begin{eqnarray}
\label{supX}
 F_{\rm
s}\approx 1+100N_\alpha \delta^{-1}.
\end{eqnarray}
Since the energy of Ly$\alpha$ photons (10.2 eV) is far above the
${\rm H^{-}}$ photodissociation threshold (0.75 eV), the
photodissociation rate grows roughly as $(1+z_i)^{1.5}/(1+z)^{1.5}$,
where $z_i$ is the redshift at which the photon was emitted. In the
case of an extended period of partial ionization, $F_{\rm s}$ may be
increased by a factor of a few, possibly exceeding
$10^4\delta^{-1}$.

Since, when the IGM temperature rises above $10^4$ K,
the formation of new minihalos is suppressed, the impact of
${\rm H^{-}}$ photodissociating flux produced by X-ray conversion is relevant
only for minihalos which have formed some time ago or for halos with $T_{\rm vir} > 10^4$ K, which also rely on $\HH$ cooling to form stars.

\section{Discussion}

As shown by our calculations, $\HM$ photodissociation reduces the
formation of $\HH$ molecules by a factor of
\begin{equation}
F_{\rm s}\sim 1+10^3 k_s xf_{\rm esc}^{-1}\delta^{-1},
\end{equation}
where $k_s$ is a constant of order a few, whose value depends on the
type of radiation source and the growth history of the radiation
background. Thus, by the time a significant fraction ($\gsim 0.1$)
of the universe becomes ionized, $\HM$ photodissociation can
significantly reduce the $\HH$ formation rate in regions with
overdensities of up to a few thousands, i.e. in the interior regions
of minihalos. The equilibrium abundance of molecular hydrogen during
this stage would be determined by the balance between its formation
and destruction rates (Eqs. [\ref{fH2}] and [\ref{fdes}])
\begin{equation}
n_\HH = \frac{k_{-}\;n_{\rm H}\;n_\HM}{k_{\rm LW}},
\end{equation}
where $k_{\rm LW}$ is the $\HH$ destruction rate by the Lyman-Werner
photons. Thus a reduction of $\HM$ abundance by a factor $F_{\rm s}$
translates into the same reduction of the $\HH$ abundance and, in
minihalos, a comparable increase of the cooling time.

Indirectly, $\HM$ photodissociation may affect the cooling in the
central regions of minihalos even during the early stages of
reionization. The maximum density that gas can reach in the core
region of a minihalo is limited by the amount of entropy it is able
to radiate away during collapse. The lower density gas prevalent
during the early collapse phase would be susceptible to $\HM$
dissociation from even a relatively low intensity $\HM$ dissociating
flux, and the resulting lowered $\HH$ abundance would limit its
ability to radiate away entropy via $\HH$ cooling. Furthermore, the
density and $\HH$ abundance at the center depend on the conditions
in the low density outer regions, through their contributions to
both the total pressure and the self-shielding ability of the halo.
We plan to investigate these effects further with numerical
radiation-hydrodynamic simulations in the future.

\acknowledgments
LC thanks the McDonald Observatory for the W.J.
McDonald Fellowship. MK gratefully acknowledges support from the
Institute for Advanced Study. This work was partially
supported by NASA Astrophysical Theory Program grants
NAG5-10825 and NNG04G177G to P. R. S.

\end{document}